# Initial Burst of Disruptive Efforts Ensuring Scientific Career Viability

Shuang Zhang[1], Feifan Liu[1], Haoxiang Xia[1,2*]

[1] Institute of Systems Engineering, Dalian University of Technology, Dalian 116024 China

[2] Institute for Advanced Intelligence, Dalian University of Technology, Dalian 116024 China

[*] Correspondence to: hxxia@dlut.edu.cn

**Despite persistent efforts to understand the dynamics of creativity of scientists over careers in terms of productivity, impact, and prize, little is known about the dynamics of scientists' disruptive efforts that affect individual academic careers and drive scientific advance. Drawing on millions of data over six decades and across nineteen disciplines, associating the publication records of individual scientists with the disruption index, we systematically quantify the temporal pattern of disruptive ideas over individual scientific careers, providing a detailed understanding of the macro phenomenon of scientific stagnation from the individual perspective. We start by checking the relationship between disruption-based and citation-based publication profiles. Next, we observe the finite inequality in the disruptive productivity of scientists, diminishing gradually as the level of disruption increases. We then identify the initial burst phenomenon in disruption dynamics. It is further revealed that while early engagement in high disruption frictions away initial productivity, compared to initial advantage in productivity or impact, initial high disruption ensures more subsequent academic viability evidenced by a longer career span and relatively final higher productivity, but does not necessarily guarantee academic success throughout careers. Further analysis shows that increasing disruptive work is uncorrelated to overall productivity but negatively correlated with the overall impact. However, increasing disruptive work in the early career is associated with higher overall productivity, yet lower overall productivity in the later career. Our research underscores the urgent need for a policy shift that encourages a balance between the pursuit of disruptive efforts and the achievement of impactful outcomes.**

## Introduction

The slowed scientific advance has caused concern(Bhattacharya & Packalen, 2020; Chu & Evans, 2021; Park et al., 2023). Despite the enormous expansion in collective effort, including the growing number of scientists, the surge in publications, and the increasing collaborations (Wuchty et al., 2007), large-scale empirical investigations suggest that these increases are not fully translated into scientific advances (Milojević, 2015). In particular, with the Disruption index (Funk & Owen-Smith, 2017), a new quantitative metric characterizing how papers change networks of citations in science, it is found that newly published papers are unlikely to disrupt existing work (Park et al., 2023), especially within large scientific fields (Chu & Evans, 2021). As Moore's law implies, the slowed growth results from the offsetting of the exponential growth in the research effort and the sharp decline in research productivity over decades (Bloom et al., 2020). The burden of knowledge (Jones, 2009, 2010), myopia in scientific information foraging (Pan et al., 2018), decay of attention (Parolo et al., 2015), and existing canon barriers (Azoulay et al., 2019; Varga, 2022) are pointed out as contributing to the slower pace. Moreover, research has revealed that large (Wu et al., 2019), familiar (Zeng et al., 2021), hierarchical (Xu et al., 2022), and remote (Lin et al., 2023) teams, fuse fewer disruptive ideas

(Leahey et al., 2023).

Scientists' disruptive efforts spur scientific advances, affecting scientific direction and efficiency (Zeng et al., 2017; Fortunato et al., 2018). It is particularly important to gain a deeper understanding of the dynamics of disruptive ideas produced by individual scientists, providing insights into the aggregated slowed scientific advance at the individual level. Decades of research have revealed the dynamics of creativity over careers in terms of productivity (Dennis, 1956; Costas et al., 2010; Petersen et al., 2012; Way et al., 2017), impact (Simonton, 1986; Cronin & Meho, 2007; Sinatra et al., 2016; Zuo & Zhao, 2021), prize (Jones, 2010; Jones & Weinberg, 2011), funding (Yin et al., 2019; Zhang et al., 2024), and other bibliometric metrics (Wu et al., 2011). Creative careers are almost infinitely varied, from early starters to late bloomers (Way et al., 2017; Weinberg & Galenson, 2019), between conceptualists and experimentalists (Cronin & Meho, 2007), with early peak or flat peak (Ao et al., 2023). A variety of factors from ability (Sinatra et al., 2016), luck (Janosov et al., 2020), competition (Petersen et al., 2012), collaboration (Amjad et al., 2017; Bu et al., 2018; Sekara et al., 2018), reputation (Petersen et al., 2014), and gender (Sugimoto & Cronin, 2012), to research fields (Jones & Weinberg, 2011) affect academic career trajectories.

Despite of complex nature of creative careers, universal patterns are identified. Research on scientific genius finds the inverted-U pattern of life-cycle creativity, observing that great achievement tends to occur around the mid-30s and early 40s (Lehman, 1953; Cole, 1979; Jones, 2010; Jones & Weinberg, 2011), and productivity tends to rise to a peak then slowly declines (Dennis, 1956; Levin & Stephan, 1991; Simonton,1997). Recently, with the increased availability of data on large-scale scientific career trajectories, research has gone one step further, identifying more quantitative patterns. It is discovered individual productivity growth follows a leptokurtic "tent-shaped" distribution that is remarkably symmetric (Petersen et al., 2012). Random rule (Sinatra et al., 2016), and hot streak (Liu et al., 2018; Liu et al., 2021) patterns are successively revealed in the dynamics of individual scientific impact, indicating highly impactful papers are randomly clustered and distributed within her body of work. Considerable effort has been devoted to quantifying, modeling, and predicting individual productivity and impact and h-index dynamics.

However, according to Kuhn's (1996) theory of scientific evolution, creativity in careers involves producing breakthrough ideas that destabilize scientific progress, not just publishing more papers or attracting more attention. Unlike novelty which refers to unconventional combinations of knowledge (Uzzi et al., 2013), and the impact which reflects the popularity in the scientific community, originality and disruption emphasize the destruction of knowledge structure caused by a paper (Funk & Owen-Smith, 2017; Wu et al., 2019). It characterizes the extent to which the paper's ideas overshadow attention to the prior work and redirect subsequent related knowledge flows. The concept of disruption is similar to the ideas of "paradigm shift" (Kuhn, 1996), "risky subversion" (Bourdieu, 1975), "displacement" (Schon, 1963), and "replacement" (Feyerabend, 1970) in classic theories of scientific change. Throughout their careers, scientists face a trade-off between conversational traditions and risky disruption. Merton (1961) notes that great scientists are seldom one-hit wonders or all-hit makers. Under the essential tension (Kuhn & Epstein, 1979), they need to seek breakthroughs to accelerate collective discovery and also to undertake incremental work to ensure steady outputs.

Deeper questions subsequently arise regarding whether scientists engage in intermittent or persistent behavior to produce disruptive work, and what consequences these prolonged outputs have for their scholarly achievements.

According to the early viewpoint proposed by Planck, younger scientists are more receptive to new ideas than older scientists (Hull et al., 1978). Super's Life-Career Rainbow implies that scientists in their late careers tend to favor more conservative paths compared to the exploration stage (Super, 1980). Simonton's model (1997) suggests that scientists have to invest substantial time to discover the most creative ideas, which tend to emerge early in their careers, peak in mid-career, and then decline. Albeit with great significance in understanding the complex process of scientists' disruptive scholarship, there is still a lack of extensive large-scale quantification of temporal patterns of disruptive outputs over individual scientific careers.

The Disruption index, which characterizes the extent of originality and disruption of papers，allows us to empirically quantify the statistical patterns in the dynamics of disruption along careers. Using millions of data over six decades and across nineteen disciplines, we in this work construct disruption-based publication profiles of individual scientists, to systematically reveal the individual disruption dynamics. Firstly, we compare the temporal characteristics of citation-based and disruption-based publication profiles of scientists. Then, we quantify the heterogeneity of their disruption productivity. Next, we pinpoint the initial burst of disruption phenomenon and delve into the possible explanation by comparing the effects of initial advantages in disruption, citation, and productivity on career trajectories. We then probe into the association between increasing disruptive work and academic performance. Finally, we investigate how disruptive behavior changes over the past six decades and make discipline comparisons.

## Results

**Asynchrony in the Relationship between Citation-based and Disruption-based Paper Profiles.** We start by constructing and comparing citation-based and disruption-based publication profiles of individual scientists' careers. Figure 1a displays the citation-based publication profile of a scientist, illustrating the evolution of academic recognition or popularity of papers throughout her scientific career. The citation network structures of papers, formed by their references and citations, depict the extent to which papers have perturbed the follow-on studies (Figure 1b). The disruption-based profile of the given scientist, shown in Figure 1c, associates each paper in sequence with a disruption index, manifesting the variations in the originality or disruption of the papers over individual scientists' careers, allowing further quantification of the temporal patterns of individual disruption dynamics.

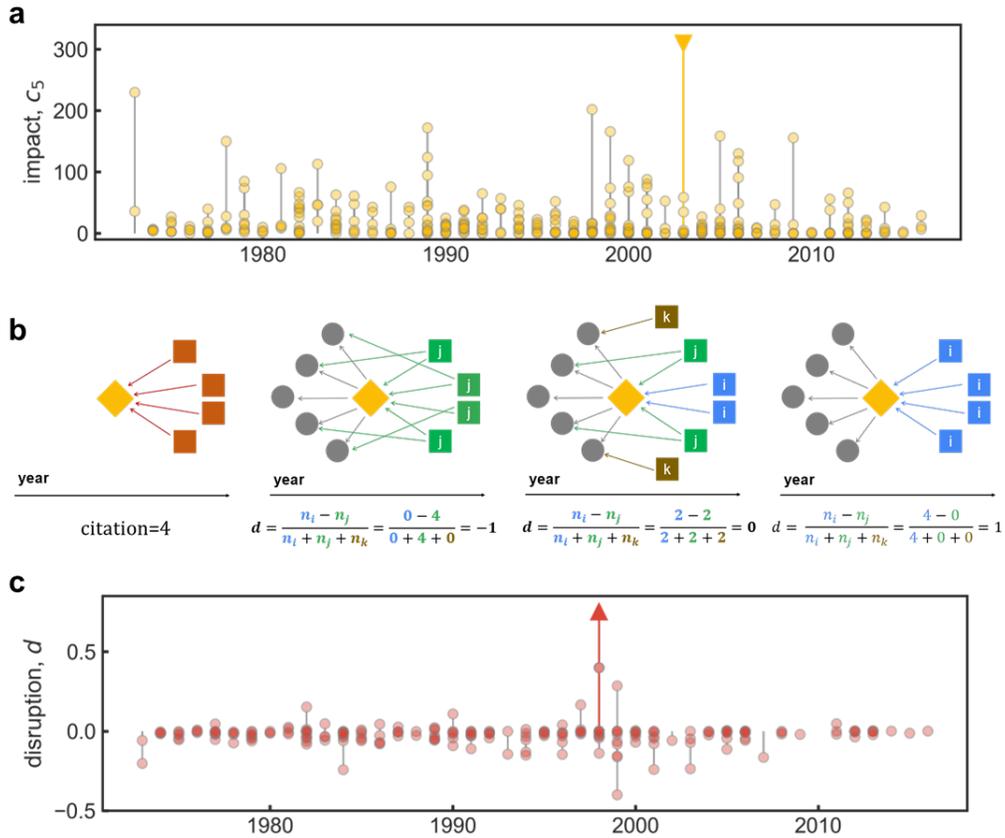

**Figure 1 | Illustration of the dynamics of disruption during individual scientists' careers**. **a**, The citation-based paper sequence of a focal scientist. The horizontal axis indicates the published years, and the height of the circle represents the publication's impact, quantified by $c_5$, the number of citations after 5 years. The highest-impact paper of the focal scientist is denoted with a yellow triangle. **b**, Demonstration of disruption index d. Three types of citation networks consisting of a particular paper (diamond), its references (circle) and its citing papers (square), correspond to the developing, neutral, and disruptive cases, respectively. **c**, The disruption-based publication sequence of a focal scientist. The height of the circle represents the publication's disruption. The highest-disruption publication of the focal scientist is denoted with a red triangle.

Notwithstanding the discrepancy in the definitions of the two metrics and the visual patterns presented in the two trajectories (Figures 1a & c), we attempt to further check the statistical level distinction between disruption-based profiles and citation-based profiles. We first test the correlation between the number of citations and disruption value per paper. In Figure 2a, both indicators are widely distributed and the curve marking the average is relatively flat. Our statistical analysis shows no significant correlation between these two indicators *(Pearson's r=.0064, p<.005)*. Next, we approach the comparison from the perspective of individual scientists by testing the Spearman rank correlation between the impact-based and the disruption-based publication series for each scientist. To be specific, we set up a controlled experiment by shuffling the disruption value and the impact of each paper in each scientist's paper series. The results in Figure 2b show the distribution of the actual rank correlation coefficients has a considerably larger variance compared to that of the shuffled experiment. About 1/3 of these coefficients are not statistically significant, and most of the significant coefficients are negative.

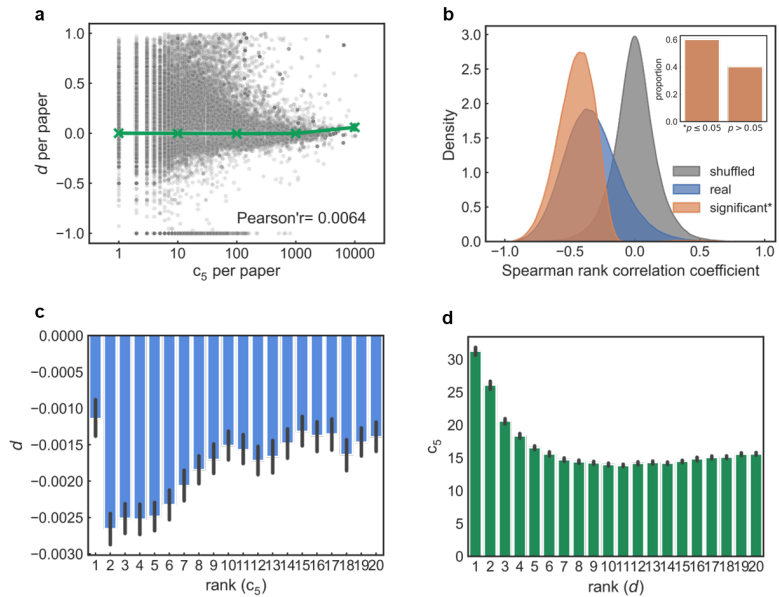

**Figure 2 | The correlation between disruption-based and citation-based paper sequences. a,** The scatter plot of the number of citations ($c_5$) versus disruption index ($d$) of papers. The curve marks the average disruption of papers with different citations. **b,** The distribution of Spearman rank correlation between disruption index ($d$) and citations ($c_5$) in individual scientists' paper sequence. The inset is the proportion of significant coefficients with a p-value less than 0.005. **c,** The average $d$ of the top 20 papers in descending order of $c_5$ in individual sequences. **d,** The average $c_5$ of the top 20 papers ranked descending by $d$ in individual sequences.

Further investigations target the typical high-impact papers and high-disruption papers in individual paper sequences. We examine the disruption (citation) of papers with the top 20 citation-rank (disruption-rank) in individual paper sequences. We observe that the average disruption shows no discernible trend with increasing citation rank (Figure 2c). However, the average number of citations decreases with increasing disruption rank up to about 10 (Figure 2d). The asymmetric correlation pattern suggests that, for individual scientists, highly disruptive papers tend to receive higher citations and be relatively more influential within their body of work. However, these influential papers may not necessarily be their most disruptive work. Appendix Figures S1&S2 present correlation results for different disciplines, showing a consistent pattern in spite of varying degrees.

The coexistence of insignificant correlations, negative correlations, and asymmetric positive correlations (Figure 2), highlights the multifaceted nature of the relationship between the dynamics of disruption and impact throughout careers. This asynchronous and nuanced interplay suggests that disruption and impact are not experienced completely consistently or inconsistently over and across scientists' careers. While impact dynamics have been widely examined previously (Sinatra et al., 2016), the evolution of disruption over careers remains scant and deserves further analysis.

**Scarcity and Inequality of Disruption in Individual Scientific Careers.** Then, we examine the variation in disruptive outputs among scientists. Figure 3a presents the heterogeneity in the distribution of *d* values of papers. As further shown in Figure 3b, the majority of papers could be considered as developing (*d < 0*), with about 25% being considered disruptive (*d > 0*).

The distribution of papers with varying levels of disruption index among scientists is then analyzed using the Gini coefficient. Our results that Gini coefficients are all greater than zero, presented in Figure 3c, find a notable inequality in the disruption outputs of scientists. Particularly, when *d* values are larger than zero, the Gini coefficient is greater than that of the number of papers. This suggests that while nearly all scientists contribute disruptive papers, the contribution is unevenly distributed among scientists, and this degree of inequality is greater than that observed in the productivity of scientists. Furthermore, in Figure 3c, we observe a decreasing trend in inequality as *d* increases. This suggests that while scientists may produce a high proportion of disruptive papers, they are less likely to have many papers with a high degree of disruption. This finite inequality is in line with the statement that great scientists are seldom one-hit wonders or all-hit makers (Merton, 1961). Scientists may have a more balanced portfolio of papers with varying degrees of disruption. Furthermore, the identified patterns that highly disruptive work is not concentrated on a few prominent scientists is distinct from well-known citation inequality (Nielsen & Andersen, 2021; Varga, 2022), implying an unusual path of producing disruptive efforts in science.

To test whether this decreasing inequality phenomenon is for the highly disruptive papers from different disciplines, we respectively analyze 19 disciplines and find similar results (Figure 3d). The respective results of all disciplines are shown in Appendix Figure S3. In summary, our analysis reveals that disruptive outputs are scarce and unequally distributed among individual scientists.

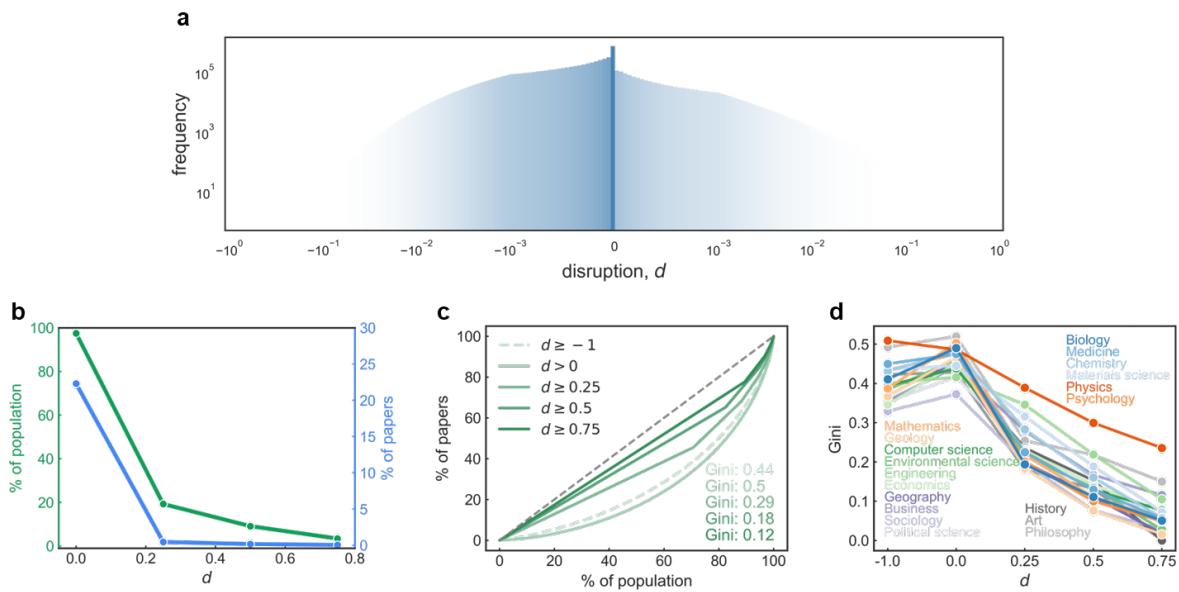

**Figure 3 | Distribution of disruptive papers in individual scientific careers. a**, The distribution of d of papers. The heavier shaded blue indicates greater frequency. **b**, Percent of the number of papers with d above a given value (blue) and percent of the number of scientists having papers with d above a given value (green). **c**, The Lorentz curves for papers with d above a given value

in the population. The grey dashed line represents the ideal equality case. The green dashed line represents the percentile of the population against cumulative papers. Large Gini coefficients correspond to a large inequality gap. **d,** The Gini coefficients at different levels of d across disciplines.

**Initial Burst of Disruption in Dynamics of Disruptive Efforts.** We have revealed the scarcity and inequality in the quantity of disruption efforts across scientist populations, as depicted in Figure 3. The next question is whether disruption efforts are distributed unevenly over individual scientists' careers. We first examine the relative position of the most disruptive paper with an individual scientist's publication sequence. Figure 4a shows a concave cumulate probability curve for $N^*/N$, indicating that most disruptive papers tend to appear earlier in the sequence of papers. This pattern is distinct from the random rule of the most-cited papers documented in Sinatra et al. (2016). For the overall average case, in Figure 4b, the average of **d** of papers presents a decreasing trend over career years. To validate the observed temporal patterns, we conduct shuffled experiments by randomly reordering the publication records of individual scientists. The comparisons between the shuffled results and the real data reveal that the concave pattern in Figure 4a and the decreasing trend in Figure 4b are significant and not due to random chance.

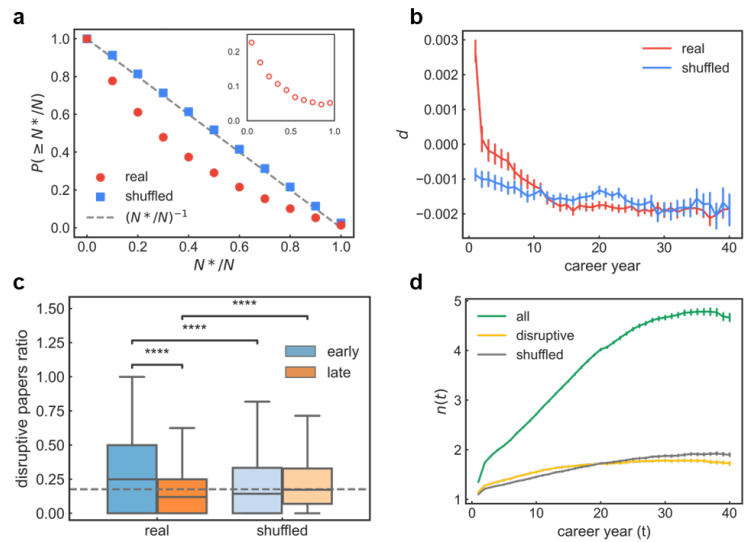

**Figure 4 | Temporal patterns of disruptive efforts during individual scientific careers. a,** Cumulative distribution of $N^*/N$ measuring the relative order ($N^*$) of the most disruptive paper in an individual sequence of N papers. A smaller $N^*/N$ indicates the most disruptive paper occurs earlier in the paper sequence. The grey line represents a uniform case. The inset shows the probability distribution of $N^*/N$. **b,** The mean d of papers published in different career years. We regard the year of the first published paper of individual scientists as their career starting years. **c,** Proportion of disruptive papers in the first and last five career years. The grey line shows the median. Papers are divided into developing (d < 0) and disruptive (d > 0) groups. The significance of difference is measured by the Mann-Whitney test. **d,** Yearly number of papers, disruptive papers for individual scientists in different career years. **a-d,** In shuffled sequences, the order of papers is randomized. **b,d,** only 40 career years are taken for plotting, due to the large fluctuations caused by the small number of careers longer than 40. Bootstrapped 95% confidence intervals are shown as error bars. ****: p ≤ .0001

To avoid extreme values of disruption confounding the observed trends, we roughly divide papers into disruptive ($d>0$) and developing papers ($d<0$). On the one hand, we examine the proportion of disruptive papers in the first and last five years of individual scientists' careers (Figure 4c). It is found that there is a higher proportion of disruptive papers in the early career stage than in the late stage, as further supported by shuffled experiment comparisons. On the other hand, we analyze the trend of yearly productivity of disruptive papers. It is observed that the yearly productivity, measured by the average number of papers published per year, increases over individual scientists' careers. While the yearly productivity of disruptive papers slowly rises and at a declining rate (Figure 4d). Compared with shuffled experiments, we find that the yearly productivity of disruptive papers is higher in the early and lower in the later career stages.

Together, our analysis reveals the initial burst of scientists' disruptive outputs. The concave cumulative probability curve in Figure 4a, the decreasing trend of the disruption index over career years in Figure 4b, and the higher proportion of disruptive papers in the early career stage compared to the late stage in Figure 4c and the lower productivity of disruptive papers in Figure 4d, all corroborate this initial high disruption rule. Disruptive efforts are not distributed evenly throughout a scientist's career but are more concentrated in the early stages, indicating that early-career scientists are more likely to produce highly disruptive work.

**Early Advantage in Disruption Ensuring Career Survival.** We attempt to investigate the underlying explanation for the initial high disruption phenomenon. One possible interpretation is selective attrition, where scientists with relatively lower levels of disruptive outputs may opt out of academia. While creating a relatively creative early-stage portfolio is a risky strategy, it can expand a scientist's research space and advance their career. Typically, scientists try out new ideas in the initial stages as input and gradually bring these ideas to fruition as they become more established in their fields. This hypothesis prompts us to probe the relationship between the extent of initial disruption and career length.

To this end, we calculate the average $d$, average number of citations, and total number of papers in the first five years of a scientist's career as proxies for their initial disruption, initial impact, and initial productivity. We also measure the career length from the scientist's first publication to their last. We compare these three initial performances of the top 10% of scientists with the longest career spans with those of the rest. As shown in Figure 5a, scientists with longer career spans produce more initial disruptive papers. However, in Figures 5b&c, we observe the opposite trend. Scientists with longer career spans initially produce fewer papers and less impactful papers, compared with those with ordinary career lengths. To rule out the possibility that results could be skewed by scientists in the long and average career groups from different decades, we match scientists to decades in which their careers start and conduct decade-specific analyses. These results consistently show that scientists with longer career lengths have significantly more initial disruptive work across decades (Figure 5a). Moreover, in certain decades, the initial productivity and initial impact of scientists with ordinary career lengths yet are observed to be higher than those of scientists in the long career groups (Figures 5b&c). These observations support the selective attrition hypothesis. The initial high disruption observed in scientists may be a result of this selective process, where those who remain in academia are those who have demonstrated their capacity for early disruptive contributions.

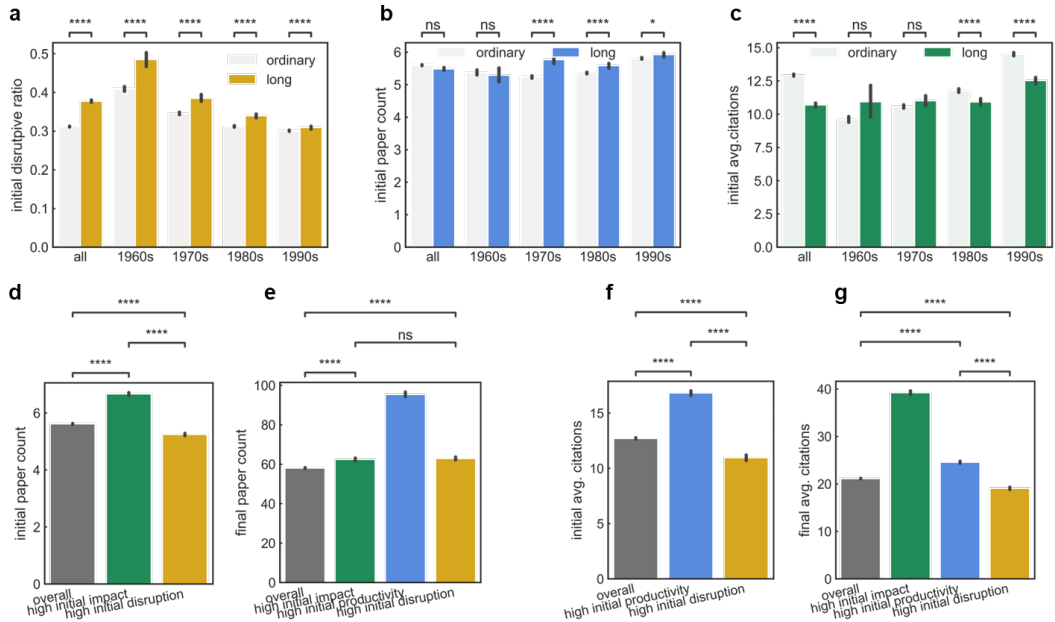

**Figure 5 | The relationship between initial advantage in disruption and career spans and final academic performance. a-c**, The disruptive papers ratio, paper count, and the average number of citations in their first five career years of scientists with longer careers (top 10%, long group) and the rest (ordinary group). Scientists are further grouped into decades in which their careers start. **d-g**, The average of the total number of papers and average citations per paper in their first five and after ten career years for different groups of scientists with the top 10% highest initial disruption, initial impact, and initial productivity, respectively. The significance of the difference is measured by the Mann-Whitney-Wilcoxon test. ****: p ≤ .0001,***: p≤ .001,**: p<= .01,*: p ≤ .05,ns:p ≥ .05

To delve deeper into the effect of the initial advantages of disruption, impact, and productivity on scientists' career paths, we compare groups of scientists respectively with the top 10% highest initial disruption, initial impact, and initial productivity, and analyze their final impact and productivity after ten career years. We find that in early careers, scientists with high initial disruption have lower productivity than the overall average (Figure 5d). However, in their final career years, these scientists achieve comparatively higher productivity, surpassing the overall average and approximately approaching the final productivity of high-initial impactful scientists (Figure 4e). This observation indicates a "friction" associated with early engagement in risky breakthrough work. It also indicates that a high initial disruption could pave the way for more future publication opportunities.

Moreover, in Figures 5f&g, both in the initial and final career stages, the average number of citations per paper of scientists with high initial disruption, is fewer than the overall average and the other two groups of scientists. However, the initial advantage in productivity and impact consistently remains throughout individual careers in Figures 5d-g. This phenomenon that the initial success translates into a sustained success is extensively studied (Krauss et al., 2023; Lee, 2023). While the initial accumulation of papers and academic recognition, coupled with the Matthew effect (Petersen et al., 2011), is necessary for career prospects, an initial burst of disruption is more of an early investment in securing career survival and laying the foundation for future opportunities, rather than ensuring long-term success.

Taken together, these results observed in Figure 5 suggest that initial prominence in disruption may contribute to longer careers and research opportunities, but it does not necessarily guarantee high academic performance in terms of productivity and impact throughout the career.

**Comparisons between productive and impactful scientists.** We further ask whether top-ranked scientists exhibit distinct dynamic patterns throughout careers. To explore this, we look at the relationship between the ratio of disruptive papers and two kinds of academic performance, productivity represented by the total number of papers, and impact measured as the average number of citations per paper. In Figures 6a&d, we observe a consistent pattern that the ratio of disruptive papers decreases as productivity and impact increase. However, when further fixing the impact and productivity of focal scientists to examine these associations, respectively, we find different stories. Scientists are classified into high, medium, and low groups based on their performance at the 90th and 10th percentiles. The inset of Figure 6a shows that in groups with fixed impact, the disruptive work ratio remains relatively stable with increasing productivity. Conversely, in the inset of Figure 6d for groups with similar productivity, the decreasing trends of disruptive work ratio with increasing impact are still observed. These observations imply that the association between disruptive efforts and academic performance is not straightforward. Increasing disruptive efforts is not positively correlated with productivity but appears to be negatively correlated with impact.

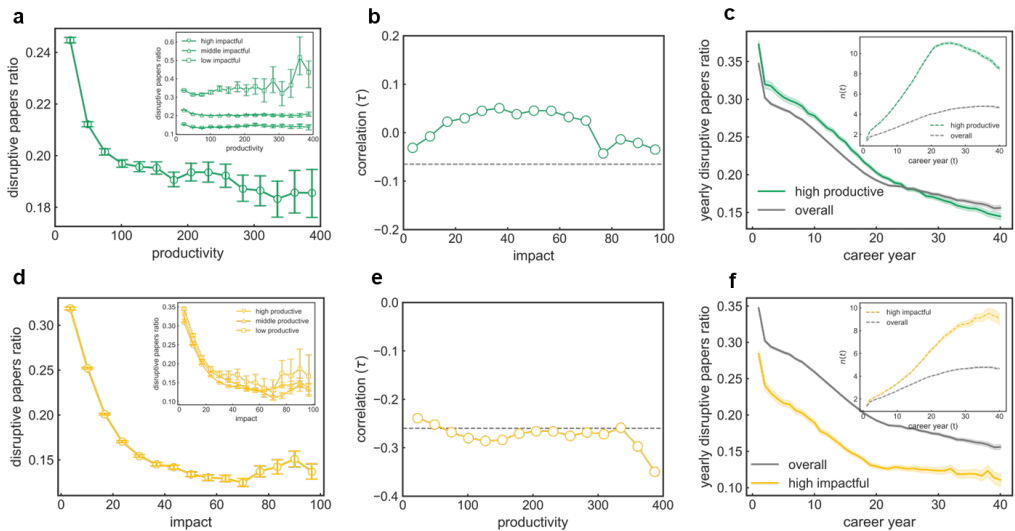

**Figure 6 | Disruptive efforts of productive and impactful scientists a**, The ratio of disruptive papers for focal scientists with different productivity. Inset of **a**, scientists are categorized into high, medium, and low impact groups by 90th percentile and 10th percentile of the average number of citations per paper. **b**, Kendall's τ correlation between the ratio of disruptive papers and productivity when fixing focal scientists' impact. The grey dashed line marks the τ coefficient in the overall case. **c**, Yearly disruptive work ratio of the 10% most productive and overall scientists over career years. Inset, the yearly number of papers produced by the 10% most productive and overall scientists over career years. **d-f**, The correlation between impact and disruptive papers ratio. Bootstrapped 95% confidence intervals are represented by error bars or colorful zones. Due to the large oscillations in the curve caused by the small data size, here we only take 400 for productivity and 100 for impact.

To further quantify the correlation, we compute Kendall's $\tau$ correlation between the disruptive work ratio and productivity or impact. In Figure 6b, although the overall Kendall's $\tau$ coefficient for productivity is slightly negative when considering all scientists together, the $\tau$ coefficients remain relatively close to zero across different levels of impact. In Figure 6c, we see the negative $\tau$ coefficient between disruptive work ratio and impact, regardless of the overall case or different productivity levels. And negative correlations are stronger for scientists with higher productivity. These results confirm that there is no significant correlation between disruptive efforts and productivity but a clear negative correlation with impact.

Next, we pay attention to the dynamics of the disruptive efforts of the highly productive scientists and highly impactful scientists (top 10%) over careers. Firstly, Figures 6c&f show that the yearly disruptive work ratio of high-productive scientists and high-impactful scientists both present decreasing trends over careers. Then, we find two opposite temporal patterns by comparing these two high-performing scientists with the overall scientists. In the early career stage, high overall productivity is associated with more disruptive work than overall, yet in the later stage, high overall productivity is associated with less disruptive work than overall (Figure 6c). By further checking the dynamics of the yearly productivity of high-productive scientists (Figure 6c, inset), we test a possible explanation for this observation: high-productive scientists' initial burst of disruption provides a bounty of "ready-to-pick fruits", evidenced by their fast-growing annual productivity in the early stages. However, as fruits are gradually consumed and become less disruptive, the annual productivity and disruptive work decrease in their later stages. Figure 6f shows that in comparison to the overall average, high-impactful scientists are involved with less disruptive work at all career stages. Concurrently, their annual productivity consistently increases over their careers. One possible explanation for this pattern is that these scientists may be more inclined to engage in easily recognizable work and away from high-risk disruptive work which typically appears in the infancy of scientific development. Additionally, high recognition attracts more resources and opportunities, which can increase productivity but may not result in as many disruptive contributions. (Figure 6f, inset).

In sum, the increasing disruptive work is uncorrelated with overall productivity but is negatively correlated with overall impact. Detailed temporal analysis reveals increasing disruptive work in the early career is associated with high overall productivity, yet with low overall productivity in the later career.

**Patterns in different decades and disciplines.** Finally, we study how the dynamics of disruption in individual careers evolve as science develops over half a century. To this end, we group focal scientists into different decades according to which period their first twenty-year careers are in. For comparability, the following analysis is conducted only on 20-year career years. The first finding is that the high-initial disruption phenomenon is observed across different decades (Figure 7a). Furthermore, it is noted that scientists nowadays make fewer disruptive contributions than those in the early years. On the one hand, Figure 7c demonstrates that papers published in the early career years of scientists who started their careers earlier are more disruptive compared to those of scientists who started their careers more recently. On the other hand, in Figure 7d, it is observed that the ratio of disruptive papers of scientists presents a decreasing trend, as science evolves. These results suggest that while the initial burst of disruption is a consistent feature across different decades, as foundational knowledge has accumulated, the increased burden of existing knowledge makes it more difficult to achieve the same level of disruption as in the past.

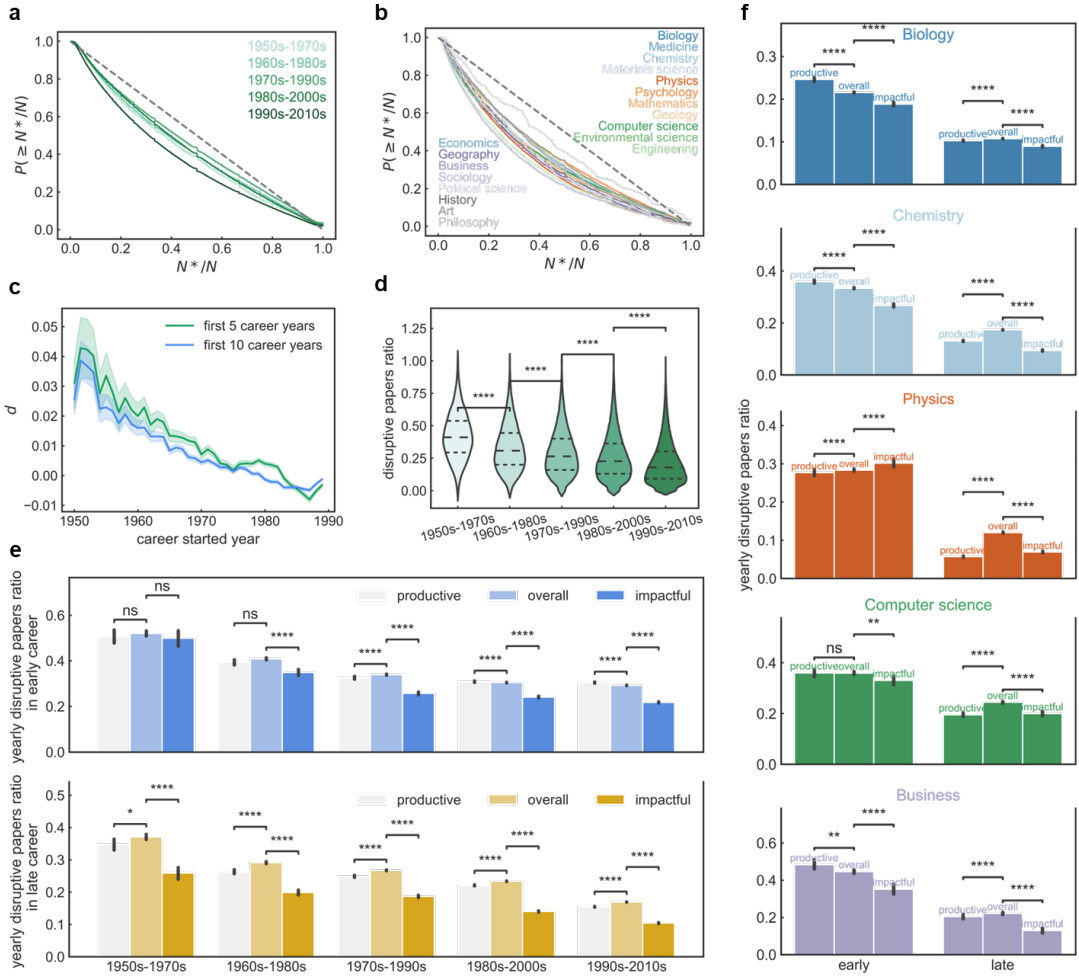

**Figure 7 | Evolution in the last century and discipline comparison. a–b,** The cumulative distribution of the relative order of the most disruptive papers in an individual publication sequence for focal scientists whose first 20-year careers in different periods (**a**), for focal scientists in different disciplines (**b**). The grey line is an ideal uniform case. **c,** The average disruption of papers published in the first five and ten career years of scientists who started their careers in different years. **d,** The disruptive work ratio for focal scientists whose first 20-year careers in different periods. **e,** The average yearly disruptive ratio in the first 5-y and last 5-y careers for the top 10% productive and the top 10% impactful scientists whose first 20-y careers in different periods. **f,** The average yearly disruptive ratio in the first 5-y and last 5-y careers for the top 10% productive and the top 10% impactful scientists in disciplines of Biology, Chemistry, Physics, Computer science, and Business. The significance of the difference is measured by the Mann-Whitney test. **** p ≤ .0001; *** p≤ .001; ** p<= .01; * p ≤ .05; ns p ≥ .05

Moreover, we compare the yearly ratio of disruptive papers in the early and late career years for high-impactful, high-productive, and overall scientists across decades. It is noted that, as science evolves, the distinction between the dynamics of disruptive efforts of high-impactful and high-productive scientists becomes more pronounced (Figure 7e). Firstly, we find the temporal pattern that high-impactful scientists produce less disruptively than overall scientists both in their early and late career years holds across decades, which is consistent with results shown in Figure 6f. Then, for high-productive scientists, the pattern that they produce less disruptive work than overall scientists in their late career years holds across decades. However, the higher than overall average disruptive work in the early career years observed in Figure 6c is not as evident in the early decades but becomes more pronounced in the late decades. These findings suggest that the distinction between high-impactful and high-productive scientists in terms of disruptive contributions is more evident as science evolves.

Furthermore, we conduct cross-disciplinary comparisons, as shown in Appendix Figures S4&S5. By and large, we observe the generalities of our findings across different disciplines. Firstly, the high-initial-disruption pattern is presented to varying degrees across disciplines (Figure 7b). This phenomenon is particularly pronounced in geography and environmental science, while it is less evident in physics and computer science. Further, we compare the temporal patterns of highly impactful and highly productive scientists across different disciplines (Figure 7f). The consistent pattern observed in Figure 6 is present in most disciplines, except for physics, where the disruption of early work by high-impact scientists is higher than that of the overall scientists. These results suggest that while the initial burst of disruption phenomenon is common across disciplines, its consequences may vary depending on the specific characteristics of a certain discipline. The distinct patterns observed in physics compared to other disciplines may be attributed to the unique nature of physics research or the way in which disruptive contributions are recognized within the discipline.

## Discussion

Given that scientists' disruptive contributions are the intrinsic drivers of scientific progress, a comprehensive, large-scale, empirical quantitative study of the dynamics of disruption over individual scientific careers is essential for understanding the efficiency of the complex scientific system. In this study, drawing on millions of papers, we construct the disruption-based profile of individual scientists' publication sequences, characterizing the dynamics of disruption over careers.

We first analyze the intricate relationship between citation-based and disruption-based profiles of individual scientists, revealing discrepancies in these trajectories over and across careers. Then, we find that rare disruptive works are unequally distributed among scientists, with inequality diminishing as disruption increases, indicating highly disruptive outputs are not concentrated on a few prominent scientists. Next, we identify the "initial burst of disruption" phenomenon, with the most disruptive papers being more prevalent in early careers. Further, by comparing initial advantages in impact and productivity, we find that though initial-high disruption compromises initial research opportunities, it enhances subsequent academic resilience, contributing to longer career lengths and relatively higher follow-on productivity. However, initial prominence in disruption does not directly fuel long-term career success. Furthermore, we note increasing disruptive work is uncorrelated to scientists' productivity but negatively uncorrelated to their impact. Further temporal analysis reveals distinct disruptive patterns between highly productive and highly impactful scientists. Finally, we find that, as science evolved, scientists nowadays make lower disruptive contributions than those in the early years, and the distinction between the dynamics of disruptive efforts of high-impactful and high-productive scientists becomes more pronounced. In addition, we observe the generality of findings across disciplines.

We further elaborate on comparisons with previous studies, explore potential explanations, and discuss policy implications. First, we observe the finite inequality characteristic of rare disruptive productivity among scientists, meaning scientists have balanced portfolios for novel and traditional research. This corresponds to the statement of Merton (1961) that great scientists are seldom one-hit wonders or all-hit makers, and inconsistent with the previously observed citation inequality that influential papers are concentrated among a small number of top scientists (Nielsen & Andersen, 2021; Varga, 2022). Secondly, our focus is on the disruptive contributions of researchers, rather than on the number of publications and citations. We find the phenomenon of the initial burst of disruption, aligns with the earlier finding Nobel prizes are received at a younger age (Jones, 2010), but is different from the previously discovered "random rule" pattern in individual citation dynamics (Sinatra et al., 2016). Yu et al (2023) recently proposed that discrepancies in empirical findings on age and impact stem from variations in size of sample selection. By analyzing textual content to differentiate impact breadth and depth, they revealed a consistent decline in creativity over a scientist's career, consistent with our findings. Packalen & Bhattacharya (2019) also find that younger scientists are more inclined to try new ideas by analyzing the novel keywords in papers. Despite our different focus and approaches, examining input ideas versus disruptive output, we arrive at a similar conclusion. Furthermore, we propose a possible explanation, selective attrition, to understand the initial burst of disruption phenomenon. We reveal distinct impacts of early advantages in disruption, impact and productivity on careers. While the initial accumulation of papers and academic recognition, compounded by the Matthew effect (Petersen et al., 2011), facilitate long-term career prospects (Krauss et al., 2023), an initial burst of disruption is more of an early investment in securing career resilience. Interestingly, we find that while the early advantage in disruption may not contribute to long-term academic performance as effectively as the other two advantages, it contributes to longer academic careers than the other two advantages. Moreover, though initial prominence in disruption may friction away early research opportunities, it guarantees more future research opportunities.

Our study reveals the temporal patterns of disruptive outputs over careers. The underlying explanation is briefly discussed here. The principle of least effort (Zipf, 1949) could be adopted to understand the initial burst of disruption, akin to our concept of selective attrition. Early-career scientists may pursue creative portfolios to expand their research space in the long term, so that as careers advance, incremental research on past breakthroughs can sustain productivity with less effort. The second is that age affects the adoption of new theories (Hull et al., 1978; Messeri, 1988). It has been empirically found older scientists tend to cite older references and collaborate with aging partners (Packalen & Bhattacharya, 2019). It is important to point out that the decreasing disruption does not indicate a decrease in cognitive abilities, but rather an increase in research complexity over careers (Liang et al., 2023). Super's Life-Career Rainbow Theory (1980) posits five stages of career development: growing, exploring, establishing, maintaining, and decreasing, where early exploratory stages favor disruptive work, maturation follows, and later stages pursue more conservative research paths. One possible explanation for the negative correlation between the impact and disruptive work of scientists is that scientists may pursue recognizable work for increasing productivity. Exploratory work typically appearing in the nascent stage of scientific development, often lacks impact and citations. The citation-based evaluation that has prevailed for nearly a century has incentivized incremental research over high-risk projects (Bhattacharya & Packalen, 2020). This trend is also reflected in the increasingly pronounced negative correlations between disruptive work and impact over decades. The possible explanation for the observation of highly productive scientists engaged in higher disruption early and in lower disruption in later careers, is initial higher disruption yields a rich harvest of achievable results, leading to rapid early productivity growth, but as these fruits are depleted, the annual productivity and disruptive work diminish in later stages. Decreasing disruption over decades may be attributed to the knowledge burden, where accumulating knowledge leads to specialization (Jones, 2009) and makes unconventional combinations of knowledge more challenging (Uzzi et al., 2013).

Our findings involved with the two major concerns of the scientific stagnation and the aging scientific workforce, provide insights into the relevant scientific policy-making. Firstly, the initial burst of disruption pattern has implications for assessing and supporting scientific careers, highlighting the importance of nurturing and supporting early-career scientists. To maintain a delicate balance between supporting young and established scientists, and between disruptive work and incremental work, funding agencies should implement distinct support mechanisms. Second, the multifaceted nature of the correlation between disruption and citation requires a holistic approach to research evaluation. Thirdly, the pursuit of breakthroughs and the desire for impact can sometimes lead to a focus on one to the detriment of the other. Our finding that early engagement in high-disruption research can benefit from career longevity and increased publication opportunities provides encouragement and confidence for undertaking potentially disruptive work.

Several research extensions could be performed. Firstly, scientific collaboration is an important feature of modern science. Further investigation in conjunction with the member roles and team age on the dynamics of the disruptive output of individuals is needed. Second, we merely discuss the hidden mechanisms for our findings. It is worthwhile to uncover the possible reasons in terms of collaboration behavior, citation behavior, and topic selection and transition. Third, scientific impact and disruption are two critical dimensions of papers. Follow-up work could explore in detail the co-evolution patterns of individual scientists' impact and disruption over careers.

## Methods

**Data.** The dataset used in this study is SciSciNet (Lin et al., 2023), a large-scale open scientific dataset constructed from the Microsoft Academic Graph (2021-12 version) (Shen et al., 2018), covering over 134 million scientific papers up to the year 2021, providing bibliography records, disambiguated authors, and fields of study. SciSciNet performs additional extensive filtering and pre-processing operations, enhancing the quality and utility of the scientific data. In data selection, the initial criterion is to include only papers published before the year 2016, for the calculation of the number of citations the paper received within 5 years ($c_5$) as an impact metric. Then, to eliminate scientists who left academia early in their careers, we adopt the criteria used by Sinatra. et al (2016), and target scientists with at least 10 papers, whose career spans at least 20 years, and who publish at least once every five years. Two points to note in this selection process. Firstly, 5-year cumulative citations are used as an impact index for papers instead of 10-year. This is done to broaden the analytical scope in terms of the number of focal scientists and the length of individual publication histories, thereby facilitating a meaningful quantification of the career dynamics. Secondly, our analysis is restricted to focal scientists' partial publication sequences valid for subsequent disruption calculation. This limitation arises from the absence of reference data or zero citations for certain papers. Therefore, we select scientists with a valid fraction of 70% or higher. Finally, we obtain 317,993 focal scientists and categorize each scientist into one of 19 disciplines based on the most frequent field label in their publication records.

**Disruption Index.** The Disruption Index ($d$) is initially designed to identify the destabilization and consolidation of patented inventions (Funk & Owen-Smith, 2017). Gradually, $d$ is introduced to measure the extent of originality or disruption of a scientific paper, characterizing whether a paper introduces new ideas eclipsing attention to the prior work on which it was built, and redirecting subsequent related citation flows (Wu et al., 2019). Impact and disruption are two critical dimensions for evaluating scientific papers, with citation metrics measuring the papers' popularity and recognition within the scientific community, while the Disruption index assesses the paper's capacity to introduce new concepts that shunt the following citation trajectories. Figure 1 illustrates the calculation of $d$ and presents three papers with distinct citation structures, despite having the same number of citations. Papers published after a given paper could be classified into three types based on their citation relationships with the given paper and its references. In this context, $n_i$ represents the number of citing papers that merely cite the given paper, $n_j$ refers to the number of citing papers that cite both the given paper and the references of the given paper, $n_k$ refers to the number of citing papers that only cite the references of the given paper. $d$ is calculated as $\frac{n_i - n_j}{n_i + n_j + n_k}$, where the difference between $n_i$ and $n_j$ is divided by all subsequent papers (sum of $n_i$, $n_j$ and $n_k$). Consequently, the value of $d$ varies between *-1* and *1*, which respectively corresponds to developing and disruptive. When $d$ is less than zero, it indicates that citing papers are more inclined to cite both the focal paper and its references ($n_i < n_j$). In this case, the focal paper is considered as developing, consolidating existing findings. When $d$ is greater than zero, it indicates that citing papers tend to only cite the focal paper and ignore its references ($n_i > n_j$). In this case, the focal paper is regarded as original and disruptive, subverting science with new ideas. When $d$ equals zero, we consider the focal paper to be neutral. The $d$ index, a better proxy to quantify the degree of disruption of a paper, has been validated extensively in previous research, including through correlation with expert assessments (Funk & Owen-Smith, 2017; Wu et al., 2019; Park et al., 2023). In this study, we use the pre-calculated and validated $d$ provided by SciSciNet (Lin et al., 2023).

# SUPPLEMENTARY INFORMATION (SI)

Initial Burst of Disruptive Efforts Ensuring Scientific Career Viability

Shuang Zhang[1#], Feifan Liu[1#], Haoxiang Xia[1,2*]

[1] Institute of Systems Engineering, Dalian University of Technology, Dalian 116024 China

[2] Institute for Advanced Intelligence, Dalian University of Technology, Dalian 116024 China

[*] Correspondence to: hxxia@dlut.edu.cn

[#] These authors contribute equally

This SI mainly includes results for different 19 disciplines.

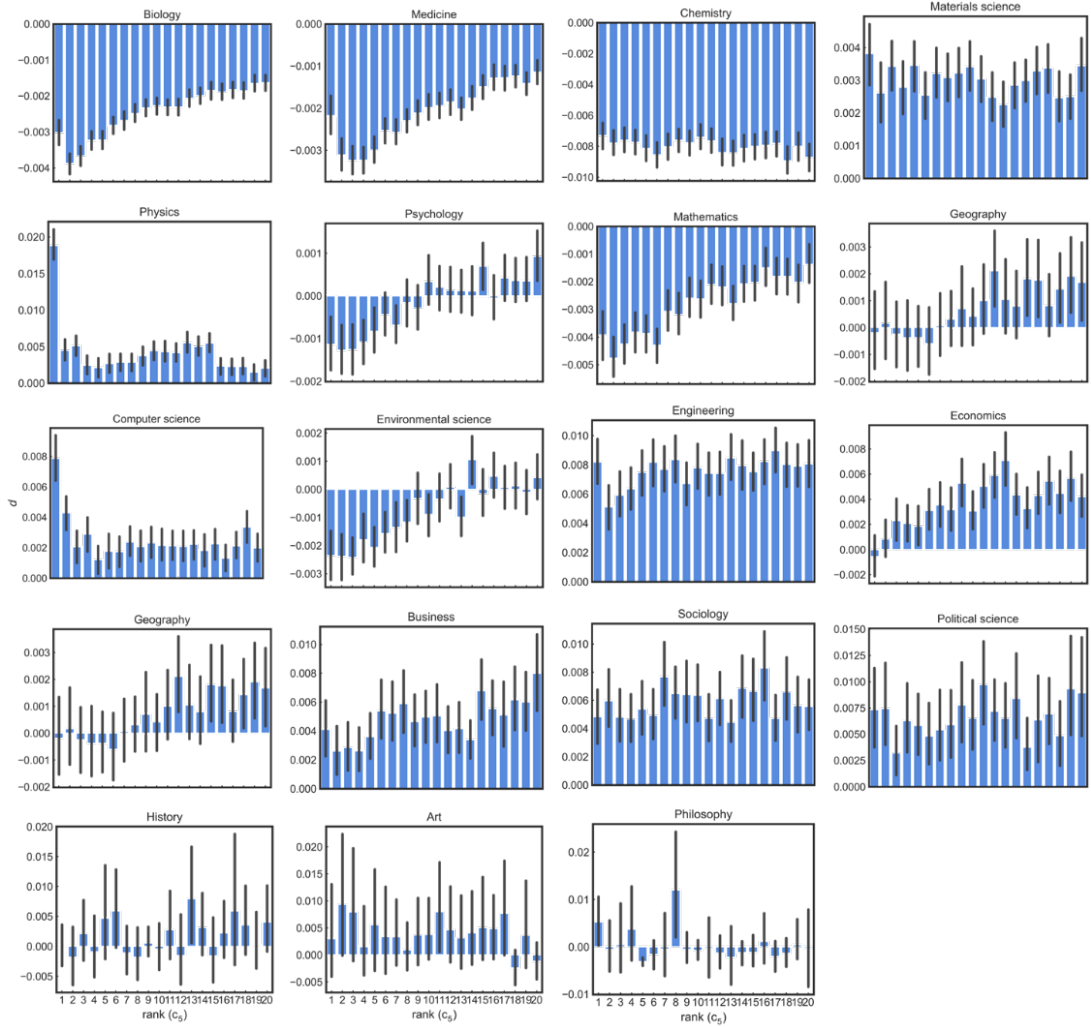

**Appendix Figure S1 | The average d of the top 20 papers in descending order of $c_5$ in individual sequences across 19 disciplines**. We examine the disruption of papers with the top 20 citation rank in individual paper sequences. It could be observed that different disciplines show different trends. And the average disruption does not exhibit a discernible trend as the citation rank increases, which is similar to the results in the main text. These results mean that these most influential papers in individual publication sequences may not necessarily be their most disruptive work.

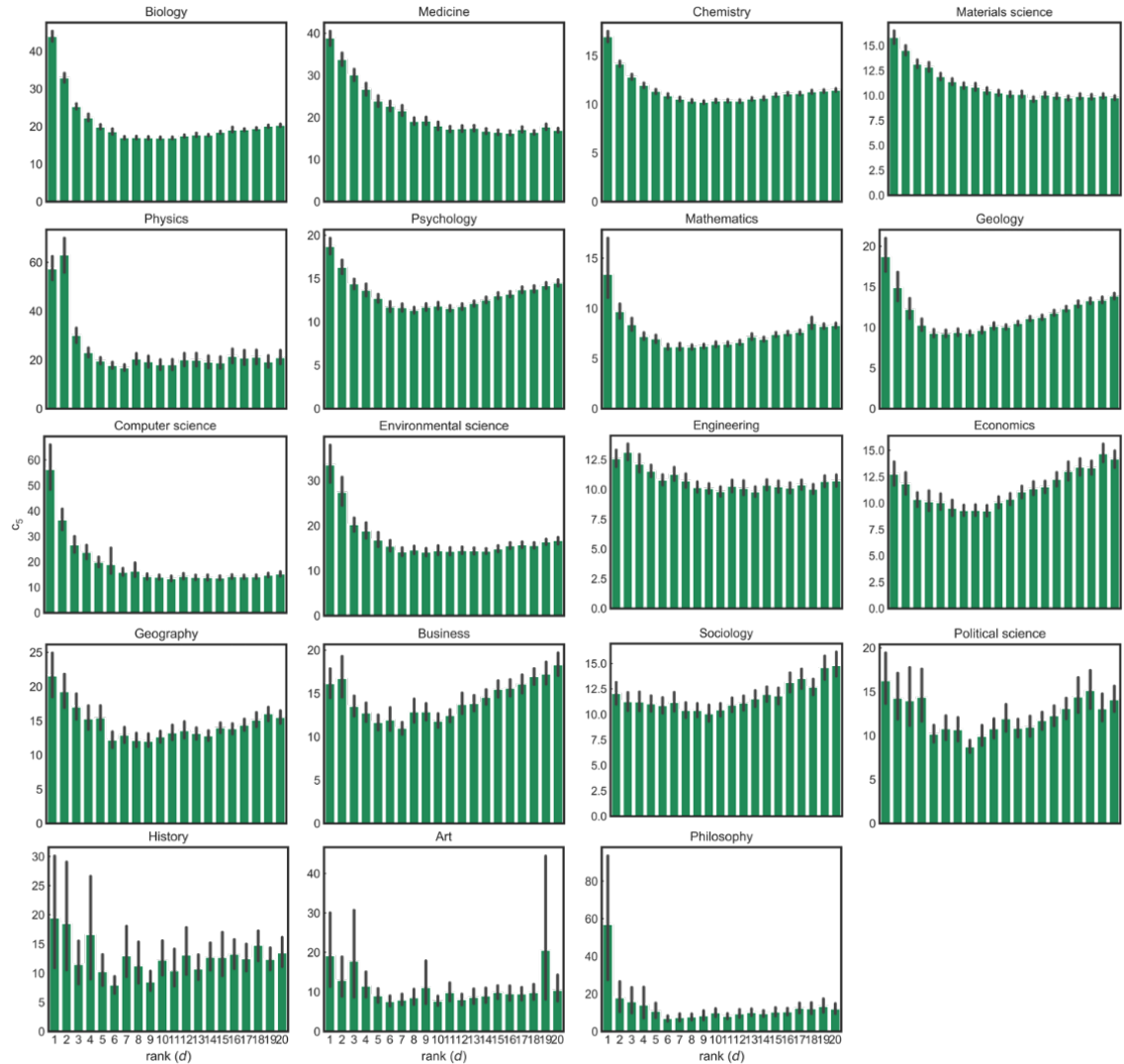

**Appendix Figure S2 | The average $c_5$ of the top 20 papers ranked descending by d in individual sequence across 19 disciplines.** We examine the average number of citations of papers with the top 20 disruption rank in individual paper sequences. It could be seen that there are similar trends presented across different disciplines. As the disruption rank increases to ten, the average number of citations presents a decreasing trend, which is similar to the observation in the main text. This result means that papers with higher disruption in individual publication sequences are likely to be their relatively most influential papers.

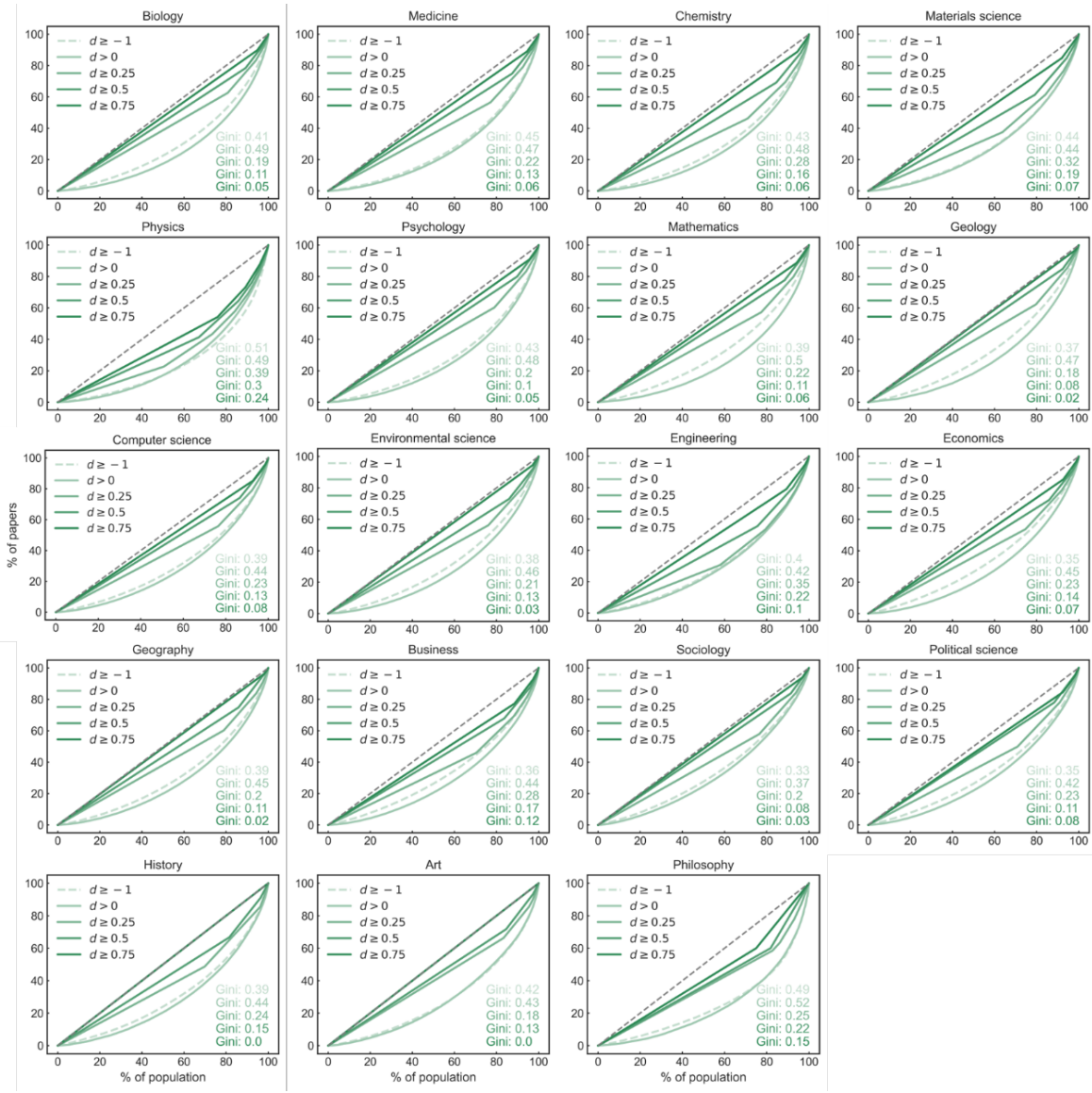

**Appendix Figure S3 | The Gini coefficients of distribution of papers with different levels of disruption index (d) across 19 disciplines.** The grey dashed line represents the ideal equality case. The green dashed line represents the percentile of the population against cumulative papers. These green solid lines represent the Lorentz curves for papers with d above a given value. Large Gini coefficients correspond to a large inequality gap. It is noted that the number of disruptive papers is unevenly distributed among scientists. And this degree of inequality is even higher than that in the productivity of scientists (dashed lines) in most disciplines, except for Materials Science, Physics, Engineering, and Art. For example, the discipline of physics presents a unique trend of first increasing and then decreasing, suggesting that the inequality of scientists' productivity is lower than that of the disruptive papers.

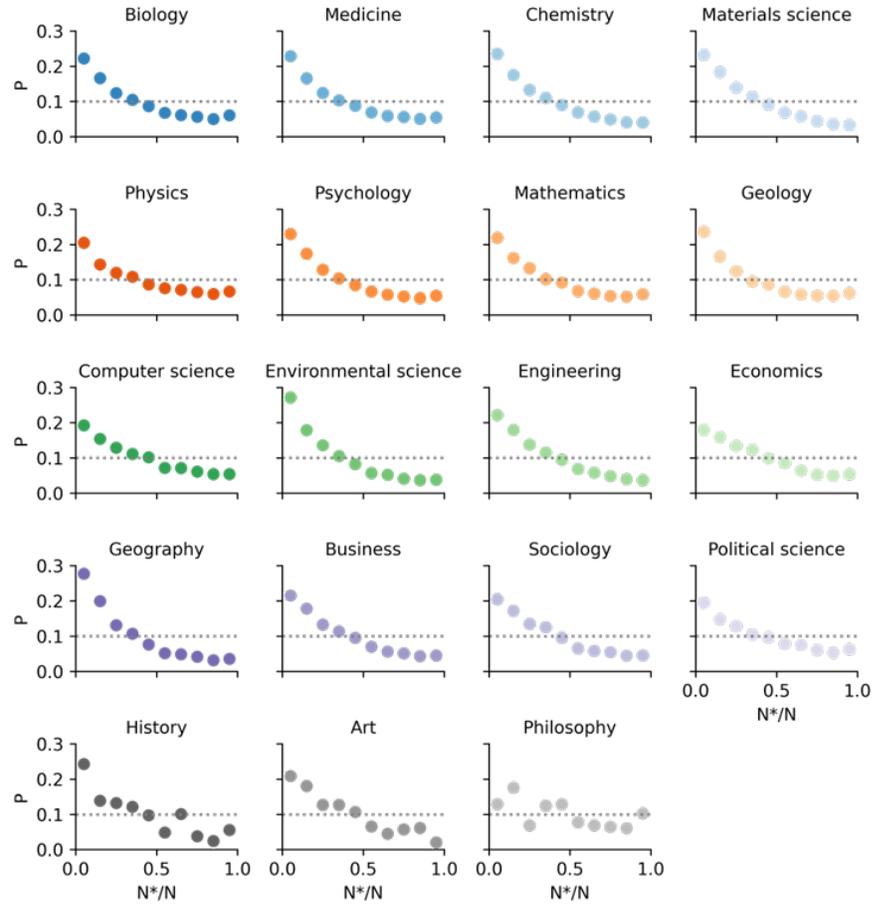

**Appendix Figure S4 | The probability distribution of N\*/N measuring the relative order (N\*) of the most disruptive paper in an individual sequence of N papers in 19 disciplines.** A smaller N\*/N indicates the most disruptive paper occurs earlier in the publication sequence. The grey dashed line represents the ideal equality case. A consistent trend of a decreasing probability curve for N\*/N is observed across disciplines, indicating that most disruptive papers tend to appear earlier in the sequence of papers.

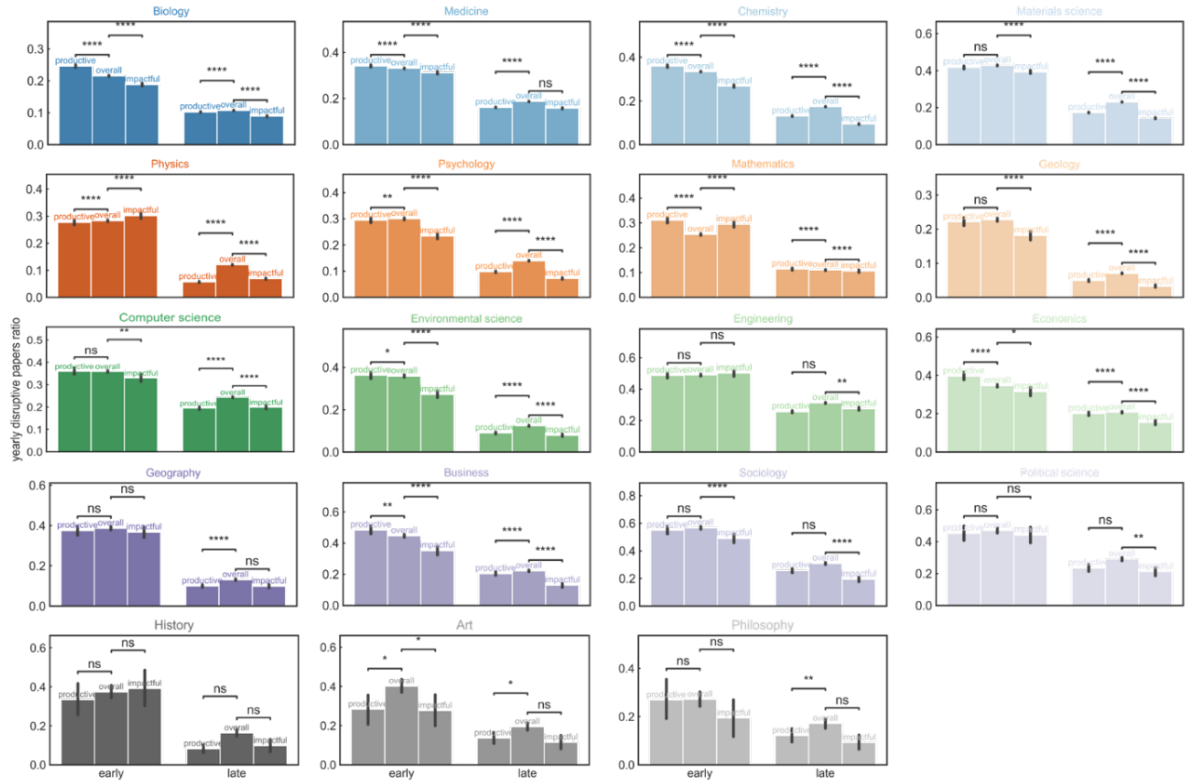

**Appendix Figure S5 | The average yearly disruptive ratio in the first 5-y and last 5-y careers for the top 10% productive and the top 10% impactful scientists in 19 disciplines.** Highly productive scientists and highly impactful scientists are respectively scientists with the top 10% of the total number of papers and the top 10% of the average number of citations per paper in their disciplines. By and large, the generalities of findings in different disciplines are observed. Except for the insignificant results, we find that high-impactful scientists produce less disruptively than overall scientists both in their early and late career years. Compared with the overall average, high-productive scientists produce higher disruptive work in the early career years and less disruptive work in the late career years. The significance of the difference is measured by the Mann-Whitney test. **** p ≤ .0001; *** p≤ .001; **p<= .01; *p ≤ .05; ns p ≥ .05.